\documentclass[prb,onecolumn,superscriptaddress]{revtex4}
\usepackage{graphicx}
\begin{document}

\title[short]{Application of elastostatic Green function tensor
technique to electrostriction in cubic, hexagonal and orthorhombic
crystals}

\date{\today}

\author{J. Hlinka}
\affiliation{Institute of Physics ASCR, Praha, Czech Republic}
\author{E. Klotins}
\affiliation{Institute of Solid State Physics, University of
Latvia, Riga, Latvia}

\begin{abstract}
The elastostatic Green function tensor approach, which was
recently used to treat electrostriction in numerical simulation of
domain structure formation in cubic ferroelectrics, is reviewed
and extended to the crystals of hexagonal and orthorhombic
symmetry. The tensorial kernels appearing in the expressions for
effective nonlocal interaction of electrostrictive origin are
derived explicitly and their physical meaning is illustrated on
simple examples. It is argued that the bilinear coupling between
the polarization gradients and elastic strain should be
systematically included in the Ginzburg-Landau free energy
expansion of electrostrictive materials.
\end{abstract}


\maketitle

\section{Introduction}

Technological advancements in ferroelectric materials have
triggered interest in the kinetics of domain pattern formation and
its description by the time dependent Ginzburg-Landau model.
Unlike the standard thermodynamic theory, valid for homogeneous
monodomain crystals, the time dependent Ginzburg-Landau model
comprises extra terms accounting for the contribution of
long-range interactions in the free energy and providing a
potential to simulate kinetics of the domain patterns, macroscopic
ferroelectric response, the impact of defects, and gives a new
insight in the piezoelectric effect. Crucial for this technique,
based on a variational derivation of the free-energy density with
respect to the polarization, is dealing with the elastic field
controlled by the inhomogeneous polarization due the
electrostrictive coupling. Indeed, the static and kinetic
properties of the domain structure are substantially influenced by
elastic strain fields associated with polarization
inhomogeneities. For example, it was shown that the
electrostrictive interactions are critical to the formation of
twin domain structure in prototype ferroelectric material $\rm
BaTiO_3$.\cite{Hu98}

The standard approach to the problem consists in elimination of
 the elastic degrees of freedom
with the help of the mechanical equilibrium conditions. This leads
to an effective interaction term $F_{\rm het}$ depending
explicitly on the order parameter (polarization $\bf P(x)$) only.
Such an effective term is then added to the Landau-Devonshire free
energy functional instead of the elastic and the electrostrictive
terms. A technical drawback of this approach is that the resulting
effective energy term is nonlinear and nonlocal. Moreover, if the
real space integration is preserved, the elastic properties of the
medium come in the formula through the anisotropic elastostatic
Green functions for which only complicated integral expressions
are known.

Nevertheless, at least for some purposes, the explicit expressions
for anisotropic elastostatic Green functions can be avoided by
expressing the interaction $F_{\rm het}$ in terms of Fourier
components of polarization. The Fourier representation is
particularly convenient in the case of modulated ferroelectrics,
where the polarization has a form of a single plane
wave\cite{Ema90}, but it is extremely useful even in the case of
3D domain structures\cite{Nam94, Hu98}. In the Fourier
representation, the elastic properties appear in the expression
for $F_{\rm het}$ through a tensorial kernel $B_{ijkl}({\bf n})$.
This tensorial kernel is 4-th order tensor angular function
 comprising all necessary information about  the electrostrictive and the
elastic properties of the material. The elastic properties of the
medium are introduced in $B_{ijkl}({\bf n})$ solely via the so
called elastic Green function tensor\cite{Sem98}, which is a much
more simple object than the (real space) elastic Green function
itself.

In some cases, the exact form of the tensorial kernel
$B_{ijkl}({\bf n})$ could be reasonably approximated by that of
elastically isotropic medium\cite{Kvi97,Sem98}. Some time ago,
however, the components of $B_{ijkl}({\bf n})$ in general cubic
crystals were derived explicitly\cite{Nam94}, and the fully
anisotropic $F_{\rm het}$ was then successfully used in realistic
2D and 3D simulations\cite{Nam94,Hu97,Hu98} of domain structure
coarsening in perovskite ferroelectrics. Objective of this work is
to the extension of this technique, which may be called
"elastostatic Green function tensor technique", to the crystals of
lower symmetry.

For the sake of clarity, we have introduced the notation and the
approach leading to the expressions for effective energy
contribution $F_{\rm het}$ in section II. The explicit expressions
known for cubic crystals\cite{Nam94} are generalized to the case
of orthorhombic and hexagonal symmetries in section III. The
section IV is devoted to the basic electrostriction (without
gradient terms). The $F_{\rm het}$ is expressed in terms of
polarization autocorrelation tensor, and the physical meaning of
$A_{ijkl}$ tensor introduced in Ref.~\onlinecite{Nam94} is
discussed in detail. Finally, the role of gradient terms in
systematic expansion of electrostrictive energy is elucidated in
the Section V.

\section{Elimination of elastic degrees of freedom }

The excess Gibbs free energy functional describing an elastically
linear ferroelectric in a general polarization and stress state
can be expressed as a sum of three terms
\begin{equation}
F = F_0\{ P_i, P_{i,j}\}+F_1\{ P_i, P_{i,j}, u_{ij}\}+F_2\{
u_{ij}\}, \label{eq1}
\end{equation}
where $P_i$,$P_{i,j}$ stands for the $i-$th cartesian component of
the polarization field and for its $j-$th spatial derivative, and
$u_{ij}$ is the $ij-$component of the (infinitesimal) strain
field.

The first part $F_0\{ P_i, P_{i,j}\}$ may be further divided into
an integral of the basic local Landau free energy density $f_{\rm
L}$, Ginzburg (gradient) energy density $f_{\rm G}$ depending on
spatial derivatives of $\bf P({\bf x })$ and the contribution of
dipole-dipole interaction $F_{\rm dip}$
\begin{equation}
F_0\{ P_i, P_{i,j}\} =F_{\rm dip} + \int (f_{\rm L} + f_{\rm G})
\,{\rm d} {\bf x}\,.
\end{equation}
The electrostrictive energy $F_1\{ P_i, P_{i,j}, u_{ij}\} $ can be
expressed as an integral over electrostriction density $f_{\rm
es}$, which is by definition linear in the elastic strain field
$u_{ij}({\bf x})$
\begin{equation}
F_1\{ P_i, P_{i,j}, u_{ij}\} = \int f_{\rm es}\,{\rm d} {\bf x},
~~f_{\rm es} = - g_{ij}u_{ij},
\end{equation}
where the leading term in expansion of coefficient $g_{ij}$
\begin{equation}
g_{ij}=g_{ij} (P_i, P_{i,j}) = q_{ijkl} P_k P_l + ...
\end{equation}
is just given by the usual electrostriction tensor $q_{ijkl}$.
 Note that Einstein summation rule is assumed in the paper.
 The last term, the proper
elastic energy, is merely a quadratic function of the elastic
strain field
\begin{equation}
F_2\{ u_{ij}\} =\int (p_{ij} u_{ij} + f_{\rm ela}) \,{\rm d} {\bf
x},~ ~ f_{\rm ela} = \frac{1}{2} C_{ijkl} u_{ij}u_{kl}\,.
\label{eqF2}
\end{equation}

In this case the total stress $\sigma_{ij}({\bf x})$ can be
divided in three contributions --- thermal stress $p_{ij}$,
describing for example the common thermal dilatation, purely
elastic stress $C_{ijkl}u_{kl} $ and the proper electrostrictive
tensile stress field $g_{ij} $ originating from coupling to the
polarization field:
\begin{equation}
 \sigma_{ij}({\bf x})=\frac{\partial f}{\partial u_{ij}}=
 p_{ij}+ C_{ijkl}u_{kl} - g_{ij}.
 \label{eqSTRESS}
\end{equation}
In static problems or in dealing with slow processes like domain
structure formation, the inhomogeneous elastic strain can often be
eliminated by means of static equilibrium conditions. The local
stress equilibrium condition $\sigma_{ij,j} =0$ can be considered
as a second order partial differential equation
 for displacement field $\bf u({\bf x})$:
\begin{equation}
 C_{ijkl}\frac{\partial^2 { u_k}}{ \partial x_j \partial x_l}=
 \frac{\partial g_{ij}}{\partial x_{j}}.
 \label{eqParcDif}
\end{equation}
This condition defines $\bf u(r)$ up to a linear form (homogeneous
strain). Let us first consider a macroscopically clamped crystal
with a large volume $V$, where the homogeneous strain is zero. In
principle, the solution to the Eq.~(\ref{eqParcDif}) satisfying
\begin{equation}
 \bar{u}_{ij} \equiv   \langle u_{ij}({\bf x}) \rangle \equiv
  \frac{1}{V}\int_{V} u_{ij}({\bf x}) \,{\rm d}
 {\bf x}=0,
 \label{eqUAV}
\end{equation}
 can be found  using the corresponding anisotropic
elastostatic Green function $G({\bf x} )$ defined by
\begin{equation}
 C_{ijkl}\frac{\partial^2  G_{km} }{ \partial x_j \partial x_l}=
 \delta ({\bf x} ) \delta_{im}\, ,
 \label{eqGF}
\end{equation}
where $\delta ({\bf x} )$ and $\delta_{ij}$ are Dirac and
Kronecker deltas.

At the same time, the formal solution for $k \neq 0$ Fourier
components follows immediately from Eq.~(\ref{eqParcDif}):
\begin{equation}
 {\bf u(k)} \equiv  \langle {\bf u(x)} \exp(-{\rm i}  {\bf k x }) \rangle   =
  \frac { {\rm i}  \Omega({\bf \hat{n}})  \cdot  g({\bf k}) \cdot {\bf k}}
  {k^2},
  \label{eqSOLU}
\end{equation}
where $\bf \hat{n}$ is the unit vector such that ${\bf k} = k {\bf
\hat{n}}$,
\begin{equation}
g_{ij}({\bf k}) \equiv \langle g_{ij}({\bf x}) \exp(-{\rm i} {\bf
k x }) \rangle
\end{equation}
are the Fourier components of the electrostrictive tensile stress
field $g_{ij}({\bf x})$ and the Green function tensor\cite{Sem98}
$\Omega({\bf \hat{n}})$ is the inverse of the Christofell
(acoustical\cite{Bur93}) tensor $\Gamma({\bf \hat{n}})$:
\begin{equation}
 (\Omega({\bf \hat{n}}) ^{-1})_{ij}=\Gamma_{ij}({\bf \hat{n}})
 \equiv  C_{ijkl}\hat{n}_k \hat{n}_l .
 \label{eqChrist}
\end{equation}
Assuming a Born-K\'arm\'an-like boundary conditions on the volume
$V$, the inverse Fourier transform provides the heterogeneous
strain field as\cite{Ema90}
\begin{equation}
   \tilde{u}_{ij}({\bf x})=\sum_{k \neq 0}  {\rm i} k_i u_j({\bf k })
   \exp({\rm i}  {\bf k x }).
   \label{eqSOLU2}
\end{equation}
The k-vectors involved in the summation form a discrete set spread
homogeneously over the whole Brillouin zone with density
$V/(2\pi)^3$. In fact, only the long wave contributions should be
essential (theory assumes smooth inhomogeneity or thick enough
domain walls) since otherwise the long wavelength elasticity
 considered in Eq.~(\ref{eqF2}) is not adequate.
By inserting the formal solution Eq.~(\ref{eqSOLU2}) back to
Eq.~(\ref{eq1}), we obtain the searched effective interaction term
$F_{\rm het}=F_1+F_2$ for a macroscopically clamped system in the
form\cite{Sem98,Nam94,Kha95}

\begin{eqnarray}
 F_{\rm het}&=&- \frac{V}{2} \sum_{k \neq 0} \sum_{ijkl} { \hat{n}_i}
g_{ij}({\bf k}) \Omega_{jk}({\bf \hat{n}}) g_{kl}({-\bf k}) {
\hat{n}_l} = \nonumber \\
&=& - \frac{V}{2} \sum_{k \neq 0} {\bf \hat{n}} \cdot g({\bf k})
\cdot \Omega({\bf \hat{n}}) \cdot g({-\bf k}) \cdot {\bf \hat{n}},
\label{eqFREL}
\end{eqnarray}
which does not depend on the elastic strain field any more.
 The
integrand (summand) of Eq.~(\ref{eqFREL}) is a bilinear form in
Fourier transformed tensor components $g_{ij}({\bf k} )$. It is
thus possible to use the Voigt abbreviated subscript notation
\cite{Nye} for $C_{ijkl}$ and for $g_{ij}$
\begin{eqnarray}
g_1= g_{11},~ g_2= g_{22}, ~g_3= g_{33}, \nonumber\\
g_4= g_{23},~g_5= g_{13},~g_6= g_{12},
\end{eqnarray}
and rewrite Eq.~(\ref{eqFREL}) in a compact form\cite{Nam94}
(hereafter the Greek indices go always from 1 to 6)
\begin{equation}
 F_{\rm het}= - \frac{V}{2} \sum_{k \neq 0}
 g_{\alpha}({\bf k})
B_{\alpha \beta}({\bf \hat{n}})
   g_{\beta}({-\bf
 k}).
 \label{eqFREL2}
\end{equation}

The nonlocal character of this effective interaction is better
apparent after returning back to the real space :
\begin{equation}
 F_{\rm het}= - \frac{1}{2} \int_V \int_V \tilde{g}_{ij}({\bf x}^{\prime})
 \frac{\partial^2 G_{kl}({\bf x}^{\prime}-{\bf x}^{\prime \prime}) }
 {\partial  x_{j} \partial x_{m}}
 \tilde{g}_{lm}({\bf x}^{\prime \prime}) \, {\rm d}{\bf x}^{\prime}{\rm d}{\bf x}^{\prime\prime}
 \label{eqFREL3}
 \end{equation}
where $ \tilde{g}_{ij} = g_{ij}~ - \langle g_{ij} \rangle $ is the
heterogenous part of electrostritive field $g_{ij}$. This
expression shows that the $B_{\alpha \beta}$ tensorial kernel is
actually a Fourier-transformed Hessian of the elastostatic Green
function $G({\bf x} )$ defined in Eq.~(\ref{eqGF}).

Finally, the total strain field under general macroscopic
equilibrium conditions reads
\begin{equation}
   u_{ij}({\bf x})=\bar{u}_{ij}+\tilde{u}_{ij}({\bf x}),
   \label{eqGEN}
\end{equation}
where $\tilde{u}_{ij}({\bf x})$ is given by Eq.~(\ref{eqSOLU2})
and $\bar{u}_{ij}$ is the homogeneous component defined by the
left hand side of Eq.~(\ref{eqUAV}).
 For example, the free sample
condition $\bar{\sigma}_{ij}=0$ leads to the equilibrium value (
see Eq.~(\ref{eqSTRESS}))
\begin{equation}
   \bar{u}_{ij}=  S_{ijkl}(\bar{g}_{kl}- p_{kl}),
\end{equation}
where $S_{ijkl}= (C^{-1})_{ijkl}$ is the matrix of elastic
compliances. Substitution of Eq.~(\ref{eqGEN}) back in the
original potential in Eq.~(\ref{eq1}) provides $F_1+F_2 = F_{\rm
het}+ F_{\rm hom}$ where
\begin{equation}
 F_{\rm hom}= - \frac{V}{2}  (\bar{g}_{\alpha}- p_{\alpha}) S_{\alpha \beta}
(\bar{g}_{\beta}-p_{\beta})
 \end{equation}
 and $F_{\rm het}$ is just the same as for the case of
 clamped crystal (Eqs.~(\ref{eqFREL}),~(\ref{eqFREL3})).

\section{Elastostatic Green function  tensor for  cubic,
hexagonal and orthorhombic crystals }

Provided that the explicit dependence of the
 electrostrictive tensile stress $g_{ij}$ on
polarization field appearing in the Eq.~(\ref{eqFREL2}) is known,
the effecive energy term $F_{\rm het}$ can be calculated from
Eq.~(\ref{eqFREL}), (\ref{eqFREL2}) or (\ref{eqFREL3}). Obviously,
in some cases\cite{Klo99} it is worth to tackle the
problem\cite{Mura} of calculation of the Green function
derivatives appearing in Eq.~(\ref{eqFREL3}) explicitly, while in
other cases it is possible to avoid
it\cite{Hu97,Hu98,Kvi97,Nam94}, and use Eq.~(\ref{eqFREL}) or
(\ref{eqFREL2}). For example, it was shown\cite{Hu97,Hu98} that
simulations of domain structure coarsening described by the time
dependent Ginzburg-Landau equations (including the above effective
interaction term $F_{\rm het}$) can be performed entirely in the
Fourier space.

 In
this paragraph, we will concentrate on properties of the elastic
Green function tensor $\Omega_{ij}({\bf \hat{n}})$ and the 6x6
matrix of the tensorial kernel $B_{\alpha \beta}({\bf \hat{n}})$
appearing in the Eq.~(\ref{eqFREL2}). In order to avoid numerical
inversion of the Christofell tensor (Eq.~(\ref{eqChrist})) at each
wave vector direction $\bf \hat{n}$, several authors have derived
explicit formulas for the cubic symmetry Green function tensor
$\Omega_{ij}({\bf \hat{n}})$. Among them, the approach of the
Ref.~\onlinecite{Nam94} is the most suitable here since it allows
generalization to the case of hexagonal and orthorhombic symmetry.
The essential step consists in expressing $\Gamma_{ij}$ as a sum
of diagonal part $d_j({\bf \hat{n}})$ and a tensorial square of a
convenient real vector ${\bf v}$ :
\begin{equation}
\Gamma_{ij}({\bf \hat{n}})= d_j({\bf \hat{n}}) \delta_{ij} + v_i
v_j.
 \label{eqChrist2}
\end{equation}
This is trivial in cubic crystals where obviously \cite{Nam94}
\begin{equation}
v_{i}= (C_{12}+C_{44})\hat{n}_i\, , \label{eqCUBIC}
\end{equation}
\begin{equation}
d_{i}=C_{44} +(C_{11}-C_{12}+C_{44})\hat{n}_i^2. \label{eqCUBIC2}
\end{equation}
The decomposition is not so straightforward for the crystals of
lower symmetry. Nevertheless, for example in the case of
orthorhombic elastic medium with
\begin{equation}
C_{23}>-C_{44} \, ,~~C_{13}>-C_{55}\, ,~~C_{12}>-C_{66}\, ,
\end{equation}
(which is a very weak assumption since practically all known
crystals have all the off-diagonal elements $C_{12}, C_{13}$ and
$C_{23}$ positive), the Christofell tensor is given by
Eq.~(\ref{eqChrist2}) with
\begin{eqnarray}
v_{1}=\hat{n}_1 \sqrt{
\frac{(C_{12}+C_{66})(C_{13}+C_{55})}{(C_{23}+C_{44})}} \, ,
\label{eqSQRT1}\nonumber \\
v_{2}=\hat{n}_2 \sqrt{
\frac{(C_{23}+C_{44})(C_{12}+C_{66})}{(C_{13}+C_{55})}} \, , \nonumber \\
v_{3}=\hat{n}_3 \sqrt{
\frac{(C_{13}+C_{55})(C_{23}+C_{44})}{(C_{12}+C_{66})}} \, ,
\label{eqSQRT}
\end{eqnarray}
\begin{eqnarray}
d_{1}&=&C_{11}\hat{n}_i^2 +C_{66}\hat{n}_2^2 +C_{55}\hat{n}_3^2
-v_1^2\, ,\nonumber \\
d_{2}&=&C_{44}\hat{n}_i^2 +C_{22}\hat{n}_2^2 +C_{44}\hat{n}_3^2
-v_2^2\, , \nonumber \\
d_{3}&=&C_{55}\hat{n}_i^2 +C_{44}\hat{n}_2^2 +C_{33}\hat{n}_3^2
-v_3^2.
\end{eqnarray}
Obviously, the above decomposition can be used also for hexagonal
crystals; it is sufficient to put $C_{55}=C_{44}$,
$C_{22}=C_{11}$, $C_{23}=C_{13}$ and $2C_{66}=C_{11}-C_{12}$.

For arbitrary crystal symmetry, once the explicit expressions for
$v_i$ and $d_i$ are known, the Green function tensor $\Omega_{ij}$
is obtained directly using the Lemma from the Appendix:
\begin{equation}
\Omega_{ij}({\bf \hat{n}})= \frac{\delta_{ij}}{d_j} - \frac{v_i
v_j}{d_i d_j}(1+ \sum_{k=1}^3 \frac{v_k^2}{d_k})^{-1}.
\label{eqMOJE}
\end{equation}
The tensorial kernel $B_{\alpha \beta}$ then reads
\begin{equation}
B_{\alpha \beta}({\bf \hat{n}})= \beta_{\alpha \beta}
-\theta_{\alpha} \theta_{\beta}(1+ \sum_{k=1}^3
\frac{v_k^2}{d_k})^{-1}\, , \label{eqBtensor}
\end{equation}
where
\begin{eqnarray}
\theta_{1}&=& \hat{n}_1 v_1/d_1\, ,~~ \theta_{2}=\hat{n}_2
v_2/d_2\, ,~~
\theta_{3}=\hat{n}_3 v_3/d_3\, ,\nonumber  \\
\theta_{4}&=& \hat{n}_2 v_3/d_3+\hat{n}_3 v_2/d_2\, , \nonumber \\
\theta_{5}&=& \hat{n}_1 v_3/d_3+\hat{n}_3 v_1/d_1\, , \nonumber \\
\theta_{6}&=& \hat{n}_2 v_1/d_1+\hat{n}_1 v_2/d_2\, ,
\end{eqnarray}
modifies the eq. (4.16) used in Ref.~\onlinecite{Nam94} for cubic
crystals and the components of the $\{\beta_{\alpha \beta}\}$
tensor
\begin{equation}
\left( \begin{array}{cccccc}
\frac{\hat{n}_1^2}{d_1}&0&0&0&\frac{\hat{n}_3\hat{n}_1}{d_1}&\frac{\hat{n}_2\hat{n}_1}{d_1}\\
0&\frac{\hat{n}_2^2}{d_2}&0&\frac{\hat{n}_3\hat{n}_2}{d_2}&0&\frac{\hat{n}_1\hat{n}_2}{d_2}\\
0&0&\frac{\hat{n}_3^2}{d_3}&\frac{\hat{n}_2\hat{n}_3}{d_3}&\frac{\hat{n}_1\hat{n}_3}{d_3}&0\\
0&\frac{\hat{n}_3\hat{n}_2}{d_2}&\frac{\hat{n}_2\hat{n}_3}{d_3}&\frac{\hat{n}_2^2}{d_3}+\frac{\hat{n}_3^2}{d_2}&\frac{\hat{n}_1\hat{n}_2}{d_3}&\frac{\hat{n}_1\hat{n}_3}{d_2}\\
\frac{\hat{n}_3\hat{n}_1}{d_1}&0&\frac{\hat{n}_1\hat{n}_3}{d_3}&\frac{\hat{n}_1\hat{n}_2}{d_3}&\frac{\hat{n}_1^2}{d_3}+\frac{\hat{n}_3^2}{d_1}&\frac{\hat{n}_3\hat{n}_2}{d_1}\\
\frac{\hat{n}_2\hat{n}_1}{d_1}&\frac{\hat{n}_1\hat{n}_2}{d_2}&0&\frac{\hat{n}_1\hat{n}_3}{d_2}&\frac{\hat{n}_2\hat{n}_3}{d_1}&\frac{\hat{n}_2^2}{d_1}+\frac{\hat{n}_1^2}{d_2}\\
\end{array} \right)
\end{equation}
as a function of $d_i$ simply coincide with those given previously
for cubic crystals in eq. (4.15) of Ref.~\onlinecite{Nam94}.

Let us note that in the rare cases when some of denominators in
Eqs.~(\ref{eqSQRT1}) would become zero\cite{Bur93} of negative,
the method works equally well, it is sufficient to modify these
equations in order to express the Christofell matrix in the form
assumed in Appendix.

\section{Basic electrostriction in crystal of arbitrary symmetry class }

In this section, we will assume that the electrostrictive tensile
stress $g_{ij}$ is a bilinear form of polarization components :
\begin{equation}
g_{ij}({\bf x}) = q_{ijkl}P_k({\bf x}) P_l({\bf x}),
\label{eqBASIC}
\end{equation}
where $q_{ijkl}$ is the usual electrostrictive tensor, symmetric
both in the first and second pair of indexes. In all crystal
symmetry classes, at least some of the components are nonzero.
Since it is also the lowest order term in non-piezoelectric
materials, most of the phenomenological models are limited just to
that term\cite{Ema90,Nam94,Hu98,Kvi97,Klo99}. In this case, it is
convenient to introduce\cite{Nam94} an autocorrelation tensor
$Y_{ij}({\bf k})$
\begin{equation}
Y_{ij}({\bf k}) \equiv \langle P_i({\bf x}) P_j({\bf x})
\exp(-{\rm i} {\bf k x }) \rangle \, , \label{eqAUTOCO}
\end{equation}
which is nothing else but convolution of corresponding Fourier
components of the polarization field
\begin{equation}
Y_{ij}({\bf k}) = \sum_{\bf k^{\prime}} P_i({\bf k^{\prime}})
P_j({\bf k -k^{\prime}}).
\end{equation}
The heterogeneous effective energy term $F_{\rm het}$ then
reads\cite{Nam94}
\begin{equation}
F_{\rm het} = -\frac{V}{2}\sum_{\bf k \neq 0}
    Y_{\alpha}({\bf k}) A_{\alpha \beta
}({\bf \hat{n}})Y_{\beta}({\bf - k }). \label{eqFHET4}
\end{equation}
where $A_{\alpha \delta }({\bf \hat{n}}) = q_{\alpha \beta}
B_{\beta \gamma }({\bf \hat{n}}) q_{\gamma \delta}$ now depends on
both elastic and electrostrictive material constants.

Let us now assume that the sample contains a single planar domain
wall perpendicular to a fixed direction ${\bf \hat{n}_0}$. Then,
we can keep only $\bf P({\bf k})$ and $\bf Y_{\alpha}({\bf k})$
with ${\bf k} \parallel {\bf \hat{n}_0}$ so that $A_{\alpha \beta
}({\bf \hat{n}}) =A_{\alpha \beta }({\bf \hat{n}_0}) $ can be
taken in front of the summation symbol in Eq.~(\ref{eqFHET4})
\begin{equation}
F_{{\rm het}} = -\frac{V}{2} A_{\alpha \beta }({\bf \hat{n}_0})
\left[ \sum_{\bf k }
    Y_{\alpha}({\bf k}) Y_{\beta}({\bf k })  -
    Y_{\alpha}(0) Y_{\beta}(0)    \right].
\label{eqFHET5}
\end{equation}
Let us further assume, for example, a $180\,^{\circ}$ domain wall
with $P_1,P_3 = 0$, so that only $Y_2({\bf k })$ contributes. The
Einstein summation in the above expression then reduces to a
single term
\begin{equation} F_{{\rm het} } = -\frac{V}{2}
A_{2 2}({\bf \hat{n}_0}) \left[ \langle P_1^4 \rangle - \langle
P_1^2 \rangle ^2
  \right]\, ,
\label{eqFHET6}
\end{equation}
where $\langle \rangle$ stands for the spatial average as defined
in Eq.~(\ref{eqUAV}).

Since the polarization field comes in Eq.~(\ref{eqFHET6}) via
spatial mean square deviation of $P_2^2$, it is apparent that the
nonzero contributions to $F_{\rm het}$ comes only from the region
in the vicinity of the domain wall. Our example thus allows to
give a clear interpretation to the $A_{\alpha \beta }({\bf
\hat{n}_0})$ tensor function. For a fixed domain wall profile, its
angular dependence defines, how the electrostrictive reduction of
domain wall energy varies with domain wall orientation ${\bf
\hat{n}_0}$, and its tensorial components distinguish various type
of domain walls according to the associated change in polarization
direction.
 Obviously, in a well-coarsened
domain pattern, where the polarization inhomogeneities are limited
to the domain wall regions, the $F_{\rm het}$ become again
effectively {\em local} functional, but depending on density, {\em
type} and {\em orientation} of domain walls.

Incommensurate structures with modulated polarization represent
another transparent case of 1D inhomogeneity where
Eq.~(\ref{eqFHET5}) holds. Actually, it was shown\cite{Ema90} that
the $F_{\rm rel}$ term is essential for explanation of the
dielectric anomalies of incommensurate ferroelectric ${\rm
NaNO_2}$. In the ideal case of uniaxial sinusoidal modulation with
$P_1,P_3 = 0$ and wave vector ${\bf k_0}
\parallel {\bf \hat{n}_0}$, there is only a single nonzero pair of
Fourier components of polarization $\{P_2({\bf k_0} ), P_2({-\bf
k_0} )\}$ so that
\begin{equation}
F_{\rm het} = -V A_{2,2}({\bf \hat{n}_0}) \left[P_1({\bf k_0} )
P_1({-\bf k_0} )\right]^2
 \, ,
\label{eqFHETSIN}
\end{equation}
As noted previously\cite{Ema90}, this expression does not depend
explicitly on modulation wave vector, and thus it does not vanish
in ${\bf k_0} \rightarrow 0$ limit. In fact, this observation is
not so surprising in the present context, since it follows from
the fact that the volume ratio between ``domain walls" and
``domains" is fixed by sinusoidal profile of the modulation.
Naturally, if one assumes that with decreasing ${\bf k_0} $ the
modulation becomes of a more and more ``rectangular" shape, the
``gap" between the energy of homogeneous and modulated
ferroelectric would vanish in ${\bf k_0} \rightarrow 0$ limit.

Finally, let us note that in the case of 1D inhomogeneity with a
fixed direction ${\bf \hat{n}_0}$, the first term on the right
hand side of Eqs.~(\ref{eqFHET5}) and~(\ref{eqFHET6}) can be
actually interpreted as a local term, merely renormalizing 4-th
order terms in the Landau-Devonshire potential $F_{\rm L}$. At the
same time , the second term, although non-local, depends on
polarization in the same way as the $F_{\rm hom}$.

\section{Gradient electrostriction }

Since the elastostatic Green function tensor technique described
in section II was developed for dealing with inhomogeneous
polarization configurations, it is natural to include in the free
energy expansion terms depending on spatial derivations of
polarization. In principle, consistent free energy expansion may
require such terms not only in the expansion of $F_0$, but also in
the expansion of $F_1$. Thus, instead of Eq.~(\ref{eqBASIC}), one
may need to assume (in a crystal with centrosymmetric paraelectric
phase):
\begin{equation}
g_{ij} = q_{ijkl}P_k P_l + r_{ijkl} \frac{\partial P_k}{\partial
x_l} +s_{ijklmn}\frac{\partial P_k}{\partial x_l} \frac{\partial
P_m}{\partial x_n} + \, ... \, . \label{eqFORMAL}
\end{equation}

Let us demonstrate the role of this gradient electrostriction
terms in the case of uniaxial ($P_1,P_3 = 0$) ferroelectric with
 orthorhombic paraelectric phase. Due to the choice of easy
polarization direction and the obvious symmetry constraints, all
nonzero terms in $g_{\alpha}({\bf x}) $ expansion up to the second
order in $P=P_2$ and $P_{,j} =
\partial P /\partial x_j$ can be easily enumerated
and conveniently expressed using Voigt notation:
\begin{eqnarray}
g_{1}&=& q_{12}P^2 +r_{12}P_{,2}+  s_{12i} P_{,i}^2\, , \nonumber \\
g_{2}&=& q_{22}P^2 +r_{22}P_{,2}+  s_{22i} P_{,i}^2\, , \nonumber \\
g_{3}&=& q_{32}P^2 +r_{32}P_{,2}+  s_{32i} P_{,i}^2\, , \nonumber \\
g_{4}&=&            r_{44}P_{,3}+ s_{442} P_{,3}P_{,2}\, , \nonumber \\
g_{5}&=&              s_{564} P_{,1}P_{,3}\, , \nonumber \\
g_{6}&=& r_{66}P_{,1}+ s_{662} P_{,1}P_{,2}.
\end{eqnarray}
The Fourier components $g_{\alpha}({\bf k}) $ then read
\begin{eqnarray}
g_{1}({\bf k})&=& (q_{12} - s_{12i} k_{i}^2 )Y({\bf k})
-{\rm i} r_{12}k_2 P({\bf k}) \, , \nonumber \\
g_{2}({\bf k})&=& (q_{22}- s_{22i} k_{i}^2) Y({\bf k})
-{\rm i}r_{22}k_{2}P({\bf k})\, , \nonumber \\
g_{3}({\bf k})&=& (q_{32}- s_{32i} k_{i}^2 )Y({\bf k})
-{\rm i}r_{32}k_{2}P({\bf k})\, , \nonumber \\
g_{4}({\bf k})&=&            -{\rm i} r_{44}k_{3}P({\bf k})- s_{442} k_{3}k_{2} Y({\bf k})\, , \nonumber \\
g_{5}({\bf k})&=&             - s_{564} k_{1}k_{3} Y({\bf k}) \, , \nonumber \\
g_{6}({\bf k})&=& -{\rm i} r_{66}k_{1}P({\bf k})- s_{662}
k_{1}k_{2} Y({\bf k}) \, ,
\end{eqnarray}
where we have used autocorrelation tensor component $Y({\bf
k})=Y_2({\bf k})$ defined previously in Eq.~(\ref{eqAUTOCO}).

The effective interaction $F_{\rm het}$ can be now evaluated from
Eq.~(\ref{eqFREL2}). Let us examine the case of 1D inhomogeneity
where only $P({\bf k})$ with ${\bf k} \parallel {\bf \hat{n}}_1$
parallel to the crystallographic axis $x_1$ is nonzero. Then, the
only nonzero components of the $B_{\alpha \beta}({\bf \hat{n}}_1)$
tensorial kernel (Eq.~(\ref{eqBtensor})) are
\begin{equation}
B_{11}({\bf \hat{n}}_1) =\frac{1}{C_{11}}\, ,~ B_{55}({\bf
\hat{n}}_1) =\frac{1}{C_{55}}\, ,~B_{66}({\bf \hat{n}}_1)
=\frac{1}{C_{66}}\, ,\label{eqTRIVIAL}
\end{equation}
and the $F_{\rm het}$ then reduces to sum of two terms
\begin{equation}
F_{\rm het\,1}= -\frac{V}{2} \sum_{k_1 \neq 0}
\frac{(q_{12}-s_{12i}k_i^2)^2 }{C_{11}}Y(k_1)Y(-k_1)\, ,
\end{equation}
and
\begin{equation}
F_{\rm het\,2}= -\frac{V}{2} \sum_{k_1 \neq 0} \frac{r_{66}^2
k_1^2}{C_{66}} P(k_1)P(-k_1). \label{eqSECOND}
\end{equation}
The first term has the same form as the expression in
Eq.~(\ref{eqFHET4}), except for the fact that the $s_{\alpha \beta
\gamma}$ coupling makes the generalized $A_{\alpha \beta}$ tensor
dependent also on the modulus of the ${\bf k}$ vector. Therefore,
the $s_{\alpha \beta \gamma}$ tensor terms in Eq.~(\ref{eqFORMAL})
can be safely neglected (unless the problem under study is
drastically sensitive to the inhomogeneity lengthscale, as for
example at lock-in phase transition in type-II incommensurate
systems\cite{Ema90}).

 The second term given by
Eq.~(\ref{eqSECOND}) can be straightforwardly transformed to
\begin{equation}
F_{\rm het\,2}= -\frac{1}{2} \frac{r_{66}^2}{C_{66}} \int_{V}
\left(\frac{\partial P({\bf x})}{\partial x_1}\right)^2 \, {\rm
d}{\bf x}, \label{eqSECOND2}
\end{equation}
so that it is apparent that this term renormalize the coefficient
of the lowest order gradient term in the ``Ginzburg part" $f_G$ of
the free-energy expansion (Eq.~(\ref{eq1})). The experimental
studies of bilinear coupling between soft mode and acoustic
branches by Brillouin and inelastic neutron scattering techniques
show that the $r_{ijkl}$ coupling term in Eq.~(\ref{eqFORMAL}) may
indeed cause an important renormalization of the Ginzburg term.
Probably, more pronounced effects are expected in crystals with a
small Ginzburg term. It is even believed that in some crystals
this gradient electrostriction compensate the Ginzburg term
completely what leads to the appearance of incommensurate
modulated structure\cite{Axe70,Dol19}. Unfortunately, in the
general 3D case, the effect of the the $r_{ijkl}$ coupling term
does not reduce to a simple renormalization of coefficients in the
Ginzburg free energy and the full anisotropy of $B_{\alpha
\beta}({\bf \hat{n}})$ tensorial kernel should be taken into
account.

\section{Conclusion}

The presented elastostatic Green function tensor technique
concerns with the simulation of the ferroelectric domain pattern
being a cutting edge problem both in theory of phase transitions
and technological applications. We have found that this technique,
applied recently to electrostriction in ferroelectrics with a
cubic paraelectric phase can be straightforwardly generalized to
hexagonal and orthorhombic crystals. The contribution of this
approach is most valuable for orthorhombic crystals since they are
far from isotropy and the closed formulas for elastostatic Green
functions are known only for a few very special limit
cases.\cite{Bur93}
 Unfortunately, the method outlined here is not very
convenient for tetragonal, trigonal and monoclinic symmetries
since the ${\bf v}$ vector used in decomposition of the
Christofell matrix would have nontrivial angular dependence. We
are not aware of any elegant method for inversion of Christofell
matrix in such cases.

The essential effect of the nonlocal, nonlinear and anisotropic
effective interaction term $F_{\rm het}$ consists in reduction of
domain wall energies in function of their orientation and the
associated change of polarization vector. This information is
conveniently contained in the tensor $A_{\alpha \beta}({\bf
\hat{n}})$ introduced
 in Eq.~(\ref{eqFHET4}). Polar diagrams of the $A_{\alpha \beta}({\bf
\hat{n}})$ tensorial components may thus be quite instructive for
understanding the behavior of a particular system.

Finally, the bilinear coupling between the polarization gradients
and elastic strain should not be overlooked in the realistic
simulations. The values of the corresponding tensorial
coefficients $r_{ijkl}$ can be determined for example with the
help Brillouin and inelastic neutron scattering techniques.

\begin{acknowledgements}
This work has been supported by fellowship grant in the European
Excellence Center of Advanced Material Research and Technology in
Riga (Contract No ICA1-CT-2000-70007) and partly by the Grant
Agency of the Czech Republic (Postdoc project 202/99/D066).
\end{acknowledgements}

\appendix
\section*{Appendix}

 Let us suppose that a finite regular real symmetrical
matrix $A$ can be written sum of a diagonal part $D$ and a real
multiple of a tensorial (dyadic) square ${\bf v} \otimes {\bf v}$
of a real vector ${\bf v}$ as
\begin{equation}
A_{ij}= (D+ \lambda {\bf v} \otimes {\bf v})_{ij}= d_j \delta_{ij}
+ \lambda v_i v_j,
 \label{eqA1}
\end{equation}
where $\lambda$ is real. Let $w_i \equiv {v_i}/{d_i}$. Then, the
matrix inverse to $A$ reads
\begin{equation}
A^{-1}= D^{-1} -\frac{\lambda {\bf w} \otimes {\bf w}}{1+ {\lambda
\bf v} \cdot {\bf w}}\, ,
 \label{eqA2}
\end{equation}
provided that the right hand side of Eq.~(\ref{eqA2}) exists. This
can be easily proven by multiplication of expression in
Eq.~(\ref{eqA1}) and Eq.~(\ref{eqA2}).

In the case of orthorhombic, hexagonal and cubic crystals, the
above result allows to find compact explicit expressions for the
corresponding elastostatic Green function tensors. For example in
Eqs.~(\ref{eqChrist2}) and ~(\ref{eqMOJE}), we have applied the
Eq.~(\ref{eqA2}) with $\lambda = 1$.

\end{document}